\documentclass[amsmath,twocolumn]{aastex702}
\usepackage{lipsum}
\usepackage{hyperref}
\usepackage{multirow}
\usepackage{xfrac}
\usepackage{CJK}

\newcommand{\ed}{2026~ED$_{138}$}
\defcitealias{holman26}{H26}
\newcommand{\Hcite}{\citetalias{holman26}}

\newcommand{\diracuw}{Department of Astronomy \& the DiRAC Institute, University of Washington, Seattle, USA}
\newcommand{\nau}{Department of Astronomy and Planetary Science, Northern Arizona University, Flagstaff, USA}
\newcommand{\usp}{Departamento de Astronomia, Instituto de Astronomia, Geof\'{\i}sica e Ci\^{e}ncias Atmosf\'{e}ricas, Universidade de S\~{a}o Paulo, 05508-090, S\~{a}o Paulo, SP, Brazil}
\newcommand{\cfa}{Center for Astrophysics $|$ Harvard \& Smithsonian, 60 Garden Street, MS 51, Cambridge, MA 02138, USA}

\shorttitle{Archival Analysis of 2026 ED$_{138}$}
\shortauthors{Murtagh et al.}
\graphicspath{{./}{figures/}}

\begin{document}

\title{A Century-Long Archival Search and Analysis of the Rubin Observatory-Discovered Dwarf Planet Candidate \ed}

\author[0000-0001-9505-1131]{Joseph Murtagh}
\altaffiliation{DiRAC Postdoctoral Fellow}
\affiliation{\diracuw}
\email[show]{murtagh@uw.edu}

\author[0000-0001-7335-1715]{Colin Orion Chandler}
\affiliation{\diracuw}
\affiliation{\nau}
\email{coc123@uw.edu}

\author[0000-0003-1996-9252]{Mario Juri\'{c}}
\affiliation{\diracuw}
\email{mjuric@uw.edu}

\author[0009-0005-9428-9590]{Ian Chow}
\affiliation{\diracuw}
\email{chowian@uw.edu}

\author[0000-0003-0743-9422]{Pedro H. Bernardinelli}
\affiliation{\usp}
\email{pedro.bernardinelli@usp.br}

\author[0009-0000-2234-9068]{Bruno M. Perroni} 
\affiliation{\usp} 
\email{bruno.perroni@usp.br}  

\author[0000-0001-6350-807X]{Ryder Strauss}
\affiliation{\diracuw}
\email{rstrau@uw.edu}

\author[0009-0004-6814-5449]{Sam Deen}
\affiliation{Independent Researcher}
\affiliation{Deep Random Survey, deeprandomsurvey.org}
\email{planetaryscience@yahoo.com}

\author[0000-0002-2179-3498]{Hugues Talbot}
\affiliation{CentraleSup\'elec, Inria, Universit\'e Paris-Saclay, France}
\email{hugues.talbot@centralesupelec.fr}

\author[0000-0002-1139-4880]{Matthew J. Holman}
\affiliation{\cfa}
\email{mholman@cfa.harvard.edu}

\author[0000-0003-4827-5049]{Kevin J. Napier}
\affiliation{\cfa}
\email{kevin.napier@sao.si.edu}


\begin{abstract}

The trans-Neptunian object (TNO) \ed\ was discovered in early data from the NSF-DOE Vera C. Rubin Observatory, and recovered in Dark Energy Camera images back to 2016. With an absolute magnitude of $H_V = 3.11 \pm 0.05$~mag, it is one of the largest known TNOs, with a diameter of 635-1590~km for geometric albedos between 0.25 and 0.04. Its position far to the south and close to the galactic plane kept it hidden in images that recorded it for decades. Here, we present archival observations of \ed\ spanning 89 years: twelve photographic plates taken between 1954 and 2000, a marginal detection on a 1937 Harvard plate, 2010-2017 VISTA near-infrared images, 2016-2020 SkyMapper images, and ten years of Asteroid Terrestrial-impact Last Alert System (ATLAS) photometry. The archival positions improve the long-term ephemeris fivefold, and we predict 15 occultations of $G \leq 16.5$~mag stars in 2027-2030, six of which will occur over populated land. The line of sight from Earth has turned by $\sim87^\circ$ since 1954, yet the plate magnitudes remain consistent with the present-day phase curve. Fitted plate by plate, they exclude oblate shapes with $c/a \leq 0.6$ for at least half of all spin-axis orientations and, with a present-day light curve amplitude below $\sim0.11-0.15$~mag, a Haumea-like Jacobi ellipsoid for $\sim90\%$ of them. We find a tentative 10.4~h light curve period ($\sim20.7$~h if double-peaked). Near-infrared colors place \ed\ in the BR (intermediate red) taxonomic class, with $J-H = 0.60\pm0.15$, redder than Haumea and without evidence of similar 1.5~$\mu$m water-ice absorption, supporting the conclusion that it is not a Haumea-family member. $z$-band photometry also limits Eris-like methane to $\lesssim25\%$ of the surface. Integrations over $\pm1$~Gyr show it sticks to the 12:7 resonance with Neptune for tens of megayears at a time, but it rarely remains permanently captured.

\end{abstract}


\section{Introduction} \label{sec:1}

The largest trans-Neptunian objects (TNOs) are few in number, but disproportionately informative about the outer solar system \citep{brown08}. Bodies several hundred kilometers or greater in diameter are expected to have relaxed towards hydrostatic equilibrium under their own gravity and are therefore dwarf-planet candidates \citep{tancredi08}. Their surfaces reveal whether volatile ices such as nitrogen, methane, and carbon monoxide have survived over the age of the solar system, which depends on their size and surface temperatures \citep{schaller07b, brown12b}. Their orbits preserve the imprint of Neptune's outward migration, which swept many of them into mean-motion resonances \citep{malhotra93, gladman12, nesvorny18}. Their intrinsic brightness also gives them a past unlike the faint bulk of the trans-Neptunian region -- several lay within the reach of decades-old photographic plates taken long before their nominal discovery. Eris, for example, was identified on a 1989 plate from the UK Schmidt Telescope at Siding Spring \citep{brown05}.

\ed\ is a recent addition to the population of large TNOs. \citet[][hereafter \Hcite]{holman26} report its discovery in early NSF-DOE Vera C. Rubin Observatory \citep[Rubin;][]{ivezic19} data, and its precovery in Dark Energy Camera \citep[DECam;][]{flaugher15} images from 2016-2024. From these data, \Hcite~show that it is one of the ten intrinsically brightest known TNOs, that its visible colors are moderately red, and that it lies near the edge of the high-inclination 12:7 mean-motion resonance with Neptune. It escaped earlier detection because it lies far south and within the galactic plane ($b \sim -12^\circ$ in 2026), regions most wide-area TNO surveys have avoided because standard moving-object pipelines struggle to overcome stellar crowding.

An object this bright has been within reach of wide-field photographic plate surveys since the 1950s, and of near-infrared and optical charge-coupled device (CCD) surveys of the southern galactic plane since 2010. Because \ed\ takes $\sim280$~yr to orbit the Sun, the viewing direction changes by only $\sim1^\circ$ per year, so a shape constraint from present-day photometry alone could take decades to accumulate. Old plates can measure the brightness from viewing geometries that will not recur for a century, which allows for shape to be tested now, such as with the photographic light curve of Pluto \citep{schaefer08}. Archival near-infrared images can also measure colors that constrain the surface composition before dedicated spectroscopy is possible. 

In this work, we present the archival record of \ed\ from 1937 to 2020 (Section~\ref{sec:2}) and describe the value it adds to the work of \Hcite: an improved orbit and predictions of stellar occultations (Section~\ref{sec:3}), a 72-yr photometric baseline that constrains the shape (Section~\ref{sec:4}), near-infrared colors that test the surface composition (Section~\ref{sec:5}), and the long-term behavior of \ed\ in the 12:7 resonance (Section~\ref{sec:6}). We summarize our conclusions in Section~\ref{sec:7}.

\section{Archival Observations} \label{sec:2}

\subsection{Photographic Plates} \label{sec:2.1}
We searched the digitized plates of the SuperCOSMOS Sky Survey \citep[SSS;][]{hambly01}, the SuperCOSMOS H$\alpha$ Survey \citep[SHS;][]{parker05}, the Space Telescope Science Institute (STScI) Digitized Sky Survey (DSS/DSS1), the Archives of Photographic PLates for Astronomical USE archive \citep[APPLAUSE;][]{enke24}, and the Digital Access to a Sky Century $@$ Harvard \citep[DASCH;][]{grindlay09}. For each plate we computed the position of \ed\ at the mid-exposure time, taken from the plate logs and checked against the recorded sidereal time. In doing so, we found that the dates listed for the SSS/SHS plates and STScI scans of the first Palomar Observatory Sky Survey \citep[POSS-I;][]{pennington93} are one day early. Known asteroids recorded on these plates appear at the positions predicted for the listed date plus one day -- \ed\ was found on these plates only after this correction was applied.

We also performed a new local plate solution for each detection plate, fitted to isolated Gaia DR3 stars \citep{gaia23} of $G=14-20$~mag around the position of \ed. For the survey plates, this is a gnomonic projection with second- or third-order distortion terms, fitted to 20-330 stars; for the 1937 Harvard plate and European Southern Observatory (ESO) 1-m plates described below, a linear solution suffices over the small area used. Because the plates predate the Gaia epoch by up to eight decades, we first propagated each reference star to the plate epoch using its Gaia proper motion. Over the 62~yr between 1954 and 2016, the reference stars around \ed\ move by a median of $\sim0.5$\arcsec, and the fastest 10\% by $>1.8$\arcsec, which can be comparable to or larger than measurement error on the plates. For plates in 1977-1984, this falls to $\sim0.3$\arcsec, and to $\sim0.1$\arcsec\ for plates from 2000. These solutions leave residuals of $\sim0.08-0.30$\arcsec\ per axis. The position uncertainties we adopt in Table~\ref{tab:plates} add the centroid uncertainty of \ed\ in quadrature, together with a 0.25\arcsec\ systematic floor for the survey plates.

\begin{deluxetable*}{llccccccccl}
\tabletypesize{\footnotesize}
\tablewidth{0pt}
\tablecaption{Pre-2010 photographic detections of \ed\label{tab:plates}}
\tablehead{
\colhead{UT mid-exposure} & \colhead{Plate} & \colhead{Stn} & \colhead{Band} & \colhead{R.A.} & \colhead{Decl.} & \colhead{$\sigma$} & \colhead{S/N} &
\colhead{Mag} & \colhead{$\Delta m$} & \colhead{Note} \\
 & & & & \colhead{(deg)} & \colhead{(deg)} & \colhead{(\arcsec)} & & \colhead{(mag)} & \colhead{(mag)} &
}
\startdata
1937-12-30 00:51 & Harvard A19892 & 074 & $B$ & 152.33377 & $+7.55478$ & 2.00 & 3.4 & \nodata & \nodata & marginal; see text \\
1954-04-03 06:10 & POSS-I E1029 & 261 & $R$ & 169.47909 & $-10.80344$ & 0.30 & 21 & $18.14\pm0.13$ & $-0.37$ &  \\
1954-04-03 06:41 & POSS-I O1029 & 261 & $B$ & 169.47871 & $-10.80337$ & 0.30 & 20 & $19.35\pm0.25$ & $-0.41$ &  \\
1977-03-20 15:53 & SERC-J 2975 & 260 & $B_J$ & 198.84398 & $-33.11241$ & 0.40 & 13 & $20.54\pm0.24$ & $+0.44$ & DSS1 scan \\
1982-03-19 06:27 & ESO-R 4699 & 262 & $R$ & 206.06374 & $-36.66041$ & 0.30 & 16 & $18.88\pm0.14$ & $-0.06$ &  \\
1984-03-09 07:23 & ESO-R 5581 & 262 & $R$ & 209.21612 & $-37.90682$ & 0.30 & 14 & $19.24\pm0.17$ & $+0.24$ &  \\
1991-03-24 16:27 & SERC-R 14248 & 260 & $R$ & 219.78829 & $-41.63029$ & 0.50 & 7 & $19.24\pm0.17$ & $+0.16$ & blended, PSF-subtracted \\
1991-04-14 05:35 & ESO 1-m ES09200a & 809 & $R$ & 219.35057 & $-41.64344$ & 0.65 & 7.6 & $18.91\pm0.51$ & $-0.13$ & bright star 7\arcsec \\
1991-05-08 03:51 & ESO 1-m ES09256a & 809 & $R$ & 218.76076 & $-41.55088$ & 0.57 & 5.0 & $19.05\pm0.44$ & $+0.04$ &  \\
1991-06-11 01:59 & ESO 1-m ES09312a & 809 & $R$ & 218.00576 & $-41.26643$ & 0.50 & 5.4 & $19.14\pm0.51$ & $+0.08$ &  \\
1993-05-16 04:39 & ESO 1-m ES10581a & 809 & $R$ & 221.77661 & $-42.36996$ & 0.74 & 3.8 & $19.40\pm0.50$ & $+0.37$ &  \\
1993-05-25 03:33 & ESO 1-m ES10606a & 809 & $R$ & 221.56060 & $-42.30415$ & 0.53 & 6.9 & $19.25\pm0.40$ & $+0.21$ & neighbor fitted \\
2000-06-07 13:06 & SHS SR 18859 & 260 & $R$ & 232.78160 & $-44.51941$ & 0.40 & 5 & $19.22\pm0.18$ & $+0.09$ &  \\
\enddata
\tablecomments{Positions are astrometric ICRS coordinates referred to Gaia DR3, as published by the Minor Planet Center; $\sigma$ is the adopted astrometric uncertainty per coordinate. S/N is the detection significance used for the astrometry (for SERC-R 14248 after subtraction of the neighboring star). Magnitudes are from forced PSF photometry calibrated to Gaia DR3 and converted to Johnson $B$ (blue plates, including $B_J$) or Cousins $R$; their uncertainties include the calibration scatter. $\Delta m$ is the observed minus predicted magnitude from the phase curve of Section~\ref{sec:4.1} ($H_V=3.11$, $\beta=0.137$~mag~deg$^{-1}$), in $V$-equivalent units. SSS, SHS and STScI POSS-I dates are corrected by $+1$~day (Section~\ref{sec:2.1}).}
\end{deluxetable*}

\begin{figure*}
\centering
\includegraphics[width=\textwidth]{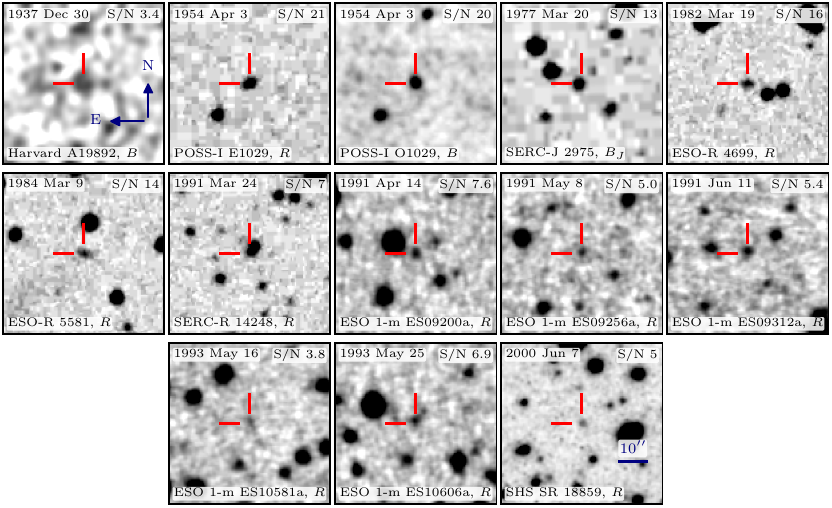}
\caption{\ed\ on thirteen photographic plates from 1937-2000. Each panel is $1'\times1'$, north up and east left, centered on the measured position (red ticks). Dates are UT of mid-exposure, corrected by $+1$~day for the SuperCOSMOS, SHS, and STScI POSS-I archive dates as described in the main text. The 1937 source (S/N 3.4) is marginal and has not been measured for photometry. That panel is smoothed with a 1.5\arcsec\ Gaussian (plate seeing is 6.3\arcsec) and shown with a stretch. On SERC-R 14248 (1991) \ed\ is blended with a $G=16.6$~mag star $\sim$2.3\arcsec\ away, and its position comes from PSF subtraction. The five ESO 1-m Schmidt plates of 1991-1993 (098-04 emulsion + OG550 filter) are APPLAUSE scans of the Hamburg plate archive.\label{fig:plates}}
\end{figure*}

\ed\ is detected on twelve plates, as shown in Figure~\ref{fig:plates}. The strongest detections are on the POSS-I red and blue plates, taken 31 minutes apart on 1954 April~3 with signal-to-noise (S/N) of 21 and 20, respectively\footnote{Independently and nearly simultaneously to our search, S. Deen was able to also measure these plates to similar precision as our work: \href{https://groups.io/g/mpml/message/42190}{https://groups.io/g/mpml/message/42190}}. It is also detected on the UK Schmidt SERC-J plate 2975 (1977), the ESO Schmidt red survey plates 4699 and 5581 (1982 and 1984, respectively), on SERC-R 14248 (1991), and on the SHS plate SR~18859 (2000). For SERC-R 14248 the detection is blended with a $G=16.6$~mag star $\sim2.3$\arcsec\ away and was measured after subtracting a model of that star. Five further detections come from deep ESO 1-m Schmidt plates of quasar-monitoring fields, taken in 1991 and 1993 with the 098-04 emulsion and an OG550 filter in 30-40~min exposures. These are fainter detections, with matched-filter S/N of 3.8-7.6. To assess their validity, we measured the same S/N at thousands of random, star-free positions on each plate; a peak as strong as the observed one occurs within our search radius only $\lesssim1.1$\% of the time. On ES09200a a bright star lies $\sim7$\arcsec\ from \ed, and on ES10606a we fitted a neighboring star simultaneously. Where plates of the same sky position from other epochs exist, we checked that the source was absent to ensure our detection was not a faint star.

The earliest plate on which \ed\ is likely recorded is Harvard plate A19892, a 60-minute exposure taken at Boyden Station on 1937 December~30. A compact, round source with S/N$\sim$3.0-3.4 lies $\sim$0.6-1.6\arcsec\ from the object's predicted position, and is absent from later images of the field in both DSS and the deeper Dark Energy Spectroscopic Instrument (DESI) Legacy Imaging Survey \citep{dey19}. A peak this strong occurs within the $\sim2.6$\arcsec\ search radius at $\lesssim1.5$\% of star-free positions on the plate. A weighted stack of the 58 deepest DASCH plates from 1900-1950, shifted along the orbit, shows a source at S/N$\sim3.2$ at the predicted position, with the signal persisting (S/N$\sim2.9$) without A19892, so \ed\ is recorded at about its predicted brightness on plates of this era. Its flux is, however, likely dominated by noise, so we treat this position as marginal and report no magnitude. Using it in the orbit fit with an adopted 2\arcsec\ uncertainty yields a residual of $\sim0.6$\arcsec, and it changes our orbital uncertainties by $\le1\%$; none of our following results depend on it. Plates before 1937 are too shallow for further detections, as \ed\ would have had a brightness of $B\sim19.6$~mag, $\sim1.5$~mag below the limit of the deepest DASCH plates.

\subsection{CCD Surveys} \label{sec:2.2}
The Visible and Infrared Survey Telescope for Astronomy (VISTA) Variables in the V\'ia L\'actea survey \citep[VVV;][]{minniti10} repeatedly imaged the southern galactic plane with the VIRCAM camera \citep{sutherland15}. We identified \ed\ in 21 VISTA images from 2010-2017, most of them from VVV. Because the field is extremely crowded, we measured $J$, $H$, and $K_s$ magnitudes on difference images, subtracting templates of the same field taken when \ed\ was elsewhere, and measured $Z$ and $Y$ magnitudes on single frames. We calibrated near-infrared magnitudes against Two Micron All Sky Survey \citep[2MASS;][]{skrutskie_two_2006} stars transformed to the VISTA system with relations from \citet{gonzalezfernandez18}. We also measured \ed\ on template-subtracted SkyMapper \citep{wolf18} images from five epochs in 2016-2020. \Hcite\ note SkyMapper candidates and 26 Asteroid Terrestrial-impact Last Alert System \citep[ATLAS;][]{tonry18} detections returned by their archival search, but did not use them -- as a result, the SkyMapper and VISTA positions, and the ATLAS photometry (Section~\ref{sec:4}), are presented here for the first time.

The VLT Survey Telescope (VST) OmegaCAM camera observed \ed\ on 2012 May~17 for the VST Photometric H$\alpha$ Survey of the Southern Galactic Plane and Bulge \citep[VPHAS+;][]{drew14}, filling the gap between the VISTA and DECam data. We detect it in the $u$, $g$, and $r$ bands with S/N$\sim10-48$, and it lies within $\sim0.04$\arcsec\ of the nominal orbit fitted without it. We calibrated the $g$ and $r$ images with aperture photometry against the synthetic Sloan Digital Sky Survey (SDSS) magnitudes derived from Gaia spectra \citep{gaia23b}, using a color term evaluated at the color of \ed. The resulting $g = 20.14 \pm 0.04$~mag and $r = 19.51 \pm 0.03$~mag agree with the phase curve of Section~\ref{sec:4.1} to $\sim0.01$~mag, and the $g-r = 0.62 \pm 0.04$~mag color equals the DECam and Rubin value measured by \Hcite~a decade later.

A search of the Canadian Astronomy Data Centre (CADC) Solar System Object Information Search \citep[SSOIS;][]{gwyn12}, Comet Asteroid Telescopic Catalog Hunter (CATCH), ESO, NOIRLab, and Mikulski Archive for Space Telescopes (MAST) archives along the 1937-2026 orbit found no earlier CCD image containing \ed. Counting the DECam images of \Hcite, \ed\ was recorded on at least twelve plates and on 264 deep CCD frames from 41 nights before its discovery. All CCD astrometry we measured is shown in Table~\ref{tab:ccd}, and has been submitted to (and accepted by) the Minor Planet Center. Table~\ref{tab:atlas} shows ATLAS measurements.

\begin{deluxetable*}{lccccccc}
\tablecaption{CCD astrometry and photometry of \ed\ from VISTA, VST and SkyMapper\label{tab:ccd}}
\tablehead{
\colhead{UT mid-exposure} & \colhead{Stn} & \colhead{Band} & \colhead{R.A.} & \colhead{Decl.} & \colhead{$\sigma$} & \colhead{Mag} & \colhead{System} \\
 & & & \colhead{(deg)} & \colhead{(deg)} & \colhead{(\arcsec)} & \colhead{(mag)} &
}
\startdata
2010-03-18 08:46:14 & W91 & $H$ & 251.025682 & $-45.488001$ & 0.08 & $17.48\pm0.13$ & Vega \\
2010-03-18 08:53:02 & W91 & $K_s$ & 251.025622 & $-45.488060$ & 0.21 & $17.47\pm0.24$ & Vega \\
2010-03-18 09:02:55 & W91 & $J$ & 251.025639 & $-45.488100$ & 0.06 & $18.07\pm0.08$ & Vega \\
2010-08-02 04:33:51 & W91 & $Y$ & 248.377357 & $-45.320360$ & 0.09 & $18.85\pm0.26$ & AB \\
2011-07-01 03:53:36 & W91 & $Z$ & 250.567893 & $-45.569462$ & 0.08 & $19.13\pm0.04$ & AB \\
2012-05-17 07:05:55 & 309 & $u$ & 253.320179 & $-45.666618$ & 0.07 & $21.74\pm0.11$ & AB \\
2012-05-17 07:20:59 & 309 & $g$ & 253.319904 & $-45.666618$ & 0.05 & $20.14\pm0.04$ & AB \\
2012-05-17 07:33:22 & 309 & $r$ & 253.319679 & $-45.666624$ & 0.05 & $19.51\pm0.03$ & AB \\
2016-08-29 08:42:42 & Q55 & $i$ & 257.719877 & $-44.765073$ & 0.28 & $19.30\pm0.20$ & AB \\
2016-08-29 08:44:42 & Q55 & $z$ & \nodata & \nodata & \nodata & $19.14\pm0.32$ & AB \\
2017-07-22 03:16:04 & W91 & $J$ & 259.696562 & $-44.934081$ & 0.05 & $18.25\pm0.08$ & Vega \\
2020-05-29 13:45:01 & Q55 & $g$ & 265.600489 & $-44.538374$ & 0.24 & $20.10\pm0.23$ & AB \\
\enddata
\tablecomments{Table~\ref{tab:ccd} is published in its entirety (35 rows: 21 VISTA, 3 VST, 11 SkyMapper) in machine-readable format, which also gives the per-coordinate uncertainties, image identifiers, and the heliocentric distance, geocentric distance, and phase angle; a portion is shown here for guidance regarding its form and content. Positions are astrometric International Celestial Reference System coordinates using Gaia DR3 as a reference; $\sigma$ is the larger of the two per-coordinate uncertainties. VIRCAM $J$, $H$ and $K_s$ (W91) are measured on template-subtracted stacks and calibrated to 2MASS stars transformed to the VISTA photometric system; $Z$ and $Y$ are PSF fits on single frames calibrated to Gaia DR3 synthetic SDSS $z$ and Pan-STARRS1 \citep{chambers_pan-starrs1_2019} $y$, and on 2010 August~2 include 0.25~mag in quadrature for a star 1.3\arcsec\ away. VST (309) $g$ and $r$ are aperture photometry and $u$ PSF photometry, calibrated to Gaia DR3 synthetic SDSS magnitudes. SkyMapper (Q55) magnitudes are forced photometry on images subtracted against DECaPS2, calibrated to Gaia DR3 synthetic SDSS magnitudes; positions are given for the five images measured astrometrically. The SkyMapper images of 2019 August~5 (residual of a bright star) and the $z$ image of 2019 May~6 (S/N 0.5) are omitted.}
\end{deluxetable*}

\begin{deluxetable*}{lccccc}
\tablecaption{ATLAS forced photometry of \ed, 2017--2026\label{tab:atlas}}
\tablehead{
\colhead{MJD (UTC)} & \colhead{Site} & \colhead{Filter} & \colhead{$N$} & \colhead{Flux} & \colhead{Mag} \\
 & & & & \colhead{($\mu$Jy)} &
}
\startdata
57958.38126 & T08 & $o$ & 3 & $90.4\pm19.5$ & $19.01\pm0.23$ \\
57979.34327 & T08 & $o$ & 4 & $6.7\pm17.2$ & $>19.62$ \\
58197.61157 & T05 & $c$ & 1 & $15.4\pm31.3$ & $>18.97$ \\
58261.46586 & T05 & $o$ & 4 & $61.1\pm14.0$ & $19.43\pm0.25$ \\
58283.44895 & T08 & $o$ & 4 & $82.6\pm15.8$ & $19.11\pm0.21$ \\
\enddata
\tablecomments{Table~\ref{tab:atlas} is published in its entirety (479 rows) in machine-readable format, which also gives the statistical flux uncertainty, the time span, the heliocentric and geocentric distances, the phase angle and the ATLAS frame identifiers; a portion is shown here for guidance regarding its form and content. Each row combines $N$ consecutive frames of one ATLAS visit (same night, site and filter); the flux is the weighted mean of forced PSF photometry at the predicted position of \ed\ on template-subtracted frames, and can be negative. The quoted flux uncertainty includes the template-subtraction noise of 0.15~mag per night, which applies to all rows of the same night, site and filter. Magnitudes are given for the 316 rows with flux $\ge3\sigma$; for the others, the magnitude column gives the 3$\sigma$ limit. Fluxes and magnitudes are in the ATLAS $o$ and $c$ bands.}
\end{deluxetable*}

\section{Orbit and Stellar Occultations} \label{sec:3}

\subsection{Orbit Determination} \label{sec:3.1}
We fitted an orbit with \texttt{layup} \citep{holman26b} to all 113 astrometric observations from 1937 to 2026, combining our archival positions (except for the VST data, which we held as validation) with the DECam and Rubin astrometry of \Hcite\ and weighing each observation by its reported uncertainty. For the Rubin data these are the per-measurement uncertainties of the pipeline, whilst for DECam we used a single value of 0.055\arcsec. We list the resulting barycentric ecliptic J2000 fitted elements and equatorial Cartesian state vector in Table~\ref{tab:fit}, with the covariance matrix shown in Table~\ref{tab:orbitcorr}. The Rubin residuals are consistent with their reported uncertainties ($\chi^2_\nu = 0.99$), whilst the DECam, VISTA, and SkyMapper residuals are smaller than the adopted uncertainties ($\chi^2_\nu \sim 0.2-0.5$). The low overall $\chi^2$ therefore reflects conservative weighting of the older CCD data. We propagated the orbit with \texttt{ASSIST} \citep{holman23}, an extension of the \texttt{REBOUND} software package \citep{rein12} for numerical $N$-body integration that uses the IAS15 integrator \citep{rein15} to integrate objects in the field of the Sun, Moon, planets, and the 16 largest asteroids. As a check, JPL Horizons \citep{giorgini96} reproduces our propagation of these elements to $\le1.2$~mas, and JPL's own orbit for \ed\ differs from ours by only $\sim1-6$~mas through 2030, almost entirely along the direction of motion.

\begin{deluxetable}{lrr}
\tablecaption{Orbit of \ed\ at epoch MJD 61203.02255891 (TDB) \label{tab:fit}}
\tablehead{\colhead{Element} & \colhead{Value} & \colhead{$1\sigma$}}
\startdata
\multicolumn{3}{c}{Barycentric ecliptic J2000 elements} \\
\hline
$a$ (au)                 & 43.09088222       & $1.3\times10^{-4}$ \\
$e$                      & 0.1387406652      & $6.2\times10^{-7}$ \\
$i$ ($^\circ$)           & 24.43388620       & $6.5\times10^{-6}$ \\
$\Omega$ ($^\circ$)      & 322.60593552      & $2.1\times10^{-5}$ \\
$\omega$ ($^\circ$)      & 168.70260900      & $1.1\times10^{-3}$ \\
$M$ ($^\circ$)           & 127.57874720      & $1.2\times10^{-3}$ \\
\hline
\multicolumn{3}{c}{Barycentric equatorial (ICRF) Cartesian state} \\
\hline
$x$ (au)                 & 2.2027956269      & $2.5\times10^{-6}$ \\
$y$ (au)                 & $-$34.7070689220  & $3.2\times10^{-5}$ \\
$z$ (au)                 & $-$31.9057658337  & $3.0\times10^{-5}$ \\
$\dot{x}$ (au\,d$^{-1}$) & 0.002303644098    & $1.8\times10^{-9}$ \\
$\dot{y}$ (au\,d$^{-1}$) & $-$0.000502293329 & $3.1\times10^{-9}$ \\
$\dot{z}$ (au\,d$^{-1}$) & 0.000348184816    & $2.4\times10^{-9}$ \\
\enddata
\tablecomments{$N_{\rm obs}=113$, $\chi^2=119$ for 220 degrees of freedom. $\omega$ and $M$ are strongly anticorrelated ($\rho=-0.9994$), so their sum is determined to $0.0002^\circ$. The covariance matrix of the Cartesian state is given in Table~\ref{tab:orbitcorr}.}
\end{deluxetable}
 
\tabletypesize{\footnotesize}
\begin{deluxetable}{lrrrrr}
\tablecaption{Covariance matrix of the Cartesian state in Table~\ref{tab:fit} \label{tab:orbitcorr}}
\tablehead{ & \colhead{$x$} & \colhead{$y$} & \colhead{$z$} & \colhead{$\dot{x}$} & \colhead{$\dot{y}$}}
\startdata
$y$       & $-$0.979371 &             &             &             &          \\
$z$       & $-$0.979348 & 0.999739    &             &             &          \\
$\dot{x}$ & 0.958686    & $-$0.963454 & $-$0.963417 &             &          \\
$\dot{y}$ & $-$0.952070 & 0.972278    & 0.971623    & $-$0.922773 &          \\
$\dot{z}$ & $-$0.931162 & 0.950398    & 0.951255    & $-$0.896946 & 0.970546 \\
\enddata
\tablecomments{The covariance is $C_{jk}=\rho_{jk}\,\sigma_j\sigma_k$, with $\sigma$ from Table~\ref{tab:fit}.}
\end{deluxetable}

To quantify the archival data's contribution, we refitted nested subsets of the observations with the same weighting. Every plate position falls within $1\sigma$ of the predicted position of \ed\ fitted using the Rubin and DECam data alone. Adding the VISTA, SkyMapper, and plate data barely changes the near-term predictions, as the Rubin 2026 observations already fix the position of \ed\ for the next few years. The improvement in predicted-position precision grows with time, however, as the archival arc constrains the orbital period and eccentricity that govern long-term drift. The uncertainty of the predicted sky position shrinks fivefold over 2035-2050, from $\sim117$ to $\sim22$~mas in 2035 and from $\sim0.53$\arcsec\ to $\sim0.1$\arcsec\ in 2050. About half of this gain derives from VISTA and SkyMapper and half from the plates, whereas the faint ESO 1-m positions and the 1937 plate add $\le1\%$.

\subsection{Occultation Predictions} \label{sec:3.2}
\begin{figure*}
\centering
\includegraphics[width=\textwidth]{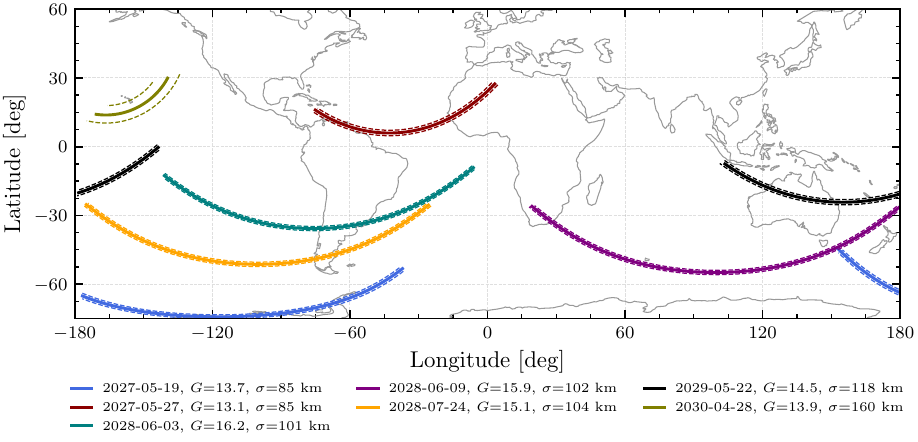}
\caption{Predicted shadow paths of \ed\ for occultations of Gaia DR3 stars with $G\le16.5$~mag in 2027-2030 whose night-side paths cross land other than Antarctica. Solid: shadow-center track where the star is more than $15^\circ$ above the horizon and the Sun more than $12^\circ$ below it. Dashed: $\pm1\sigma$ cross-track limits of the center line (given in the legend). The shadow itself is about as wide as the body ($\sim$635--1590\,km). Star positions include Gaia DR3 proper motions, and we compute ground tracks on the WGS 84 reference ellipsoid. \label{fig:occ}}
\end{figure*}

Stellar occultations provide the most direct measurements of the sizes and shapes of large TNOs, as demonstrated for Eris \citep{sicardy11}, Makemake \citep{ortiz12}, and Haumea \citep{ortiz17}. They also yield astrometry far more precise than direct imaging \citep[e.g.,][]{strauss21}, which improves the predictions of subsequent events \citep[e.g.,][]{buie21}. Occultations have revealed rings around Quaoar \citep{morgado23} and Chiron \citep{pereira25}, and can detect or place limits on tenuous atmospheres \citep[e.g.,][]{bragaribas13}. \citet{arimatsu26} recently reported evidence for a tenuous atmosphere around the $\sim500$~km TNO 2002 XV$_{93}$ from occultations, suggesting that larger bodies such as \ed\ may also retain one. The location of \ed\ near the galactic plane makes occultations frequent: the corridor along its path contains $\sim1.7\times10^5$ Gaia DR3 stars per deg$^2$ down to the Gaia limit ($G\approx21$), of which $\sim4\%$ are brighter than $G=16.5$, the approximate limit of the small portable telescopes used in many coordinated occultation campaigns \citep[e.g.,][]{buie16}. We searched Gaia DR3 for stars whose occultation shadow, propagated with the stellar proper motions, touches the Earth between 2027 and 2030, and found 385 such events. Of these, 121 occur at night (solar altitude $<-12^\circ$ at the substellar point) at solar elongations $>50^\circ$, and include stars with measured proper motions. Of these, 15 involve stars of $G\le16.5$~mag, which are bright enough for most small telescopes -- Table~\ref{tab:occ} lists these stars. Six of these have shadows that will pass over populated land (see Figure~\ref{fig:occ} -- all 121 events, including those of fainter stars accessible to larger telescopes, is available in the full machine-readable table): 2027 May~27 ($G=13.1$~mag; northern South America and West Africa), 2028 June~3 ($G=16.2$~mag; Chile, Argentina, and Uruguay), 2028 June~9 ($G=15.9$~mag; Namibia and South Africa), 2028 July~24 ($G=15.1$~mag; Patagonia), 2029 May~22 ($G=14.5$~mag; northern Australia), and 2030 April~28 ($G=13.9$~mag; Hawaii, at low altitude). A seventh, on 2027 May~19 ($G=13.7$~mag), will cross the Antarctic Peninsula and South Georgia. For a body $\sim1000$~km across, these six events will last $\sim41-80$~s.

\begin{deluxetable*}{lcccccccccl}
\tablecaption{Predicted occultations of Gaia DR3 stars with $G\le16.5$ by \ed, 2027--2030\label{tab:occ}}
\tablehead{
\colhead{UTC of closest approach} & \colhead{Gaia DR3 source} & \colhead{$G$} & \colhead{$b$} & \colhead{$v$} &
\colhead{$t_{1000}$} & \colhead{$\sigma_\perp$} & \colhead{$\sigma_\perp^{\rm sys}$} & \colhead{$\sigma_t$} & \colhead{$\sigma_t^{\rm sys}$} & \colhead{Night-side land} \\
 & & & \colhead{(km)} & \colhead{(km\,s$^{-1}$)} & \colhead{(s)} & \colhead{(km)} & \colhead{(km)} & \colhead{(s)} & \colhead{(s)} &
}
\startdata
2027-05-19 08:51:54 & 6722844039942066048 & 13.74 & 3100 & 20.0 & 50 & 85 & 190 & 5 & 11 & Antarctic Peninsula, South Georgia \\
2027-05-27 03:46:07 & 6723030029206767360 & 13.11 & 4878 & 21.9 & 46 & 85 & 190 & 4 & 10 & N South America; West Africa \\
2027-07-09 11:16:29 & 6721566544873823488 & 16.35 & 5992 & 24.8 & 40 & 87 & 193 & 4 & 8 & ocean only \\
2028-04-10 22:51:37 & 6722919390843170048 & 14.90 & 6483 & 8.8 & 114 & 162 & 286 & 17 & 30 & none at night \\
2028-05-14 17:52:19 & 6722910491670813056 & 13.73 & 2911 & 18.6 & 54 & 100 & 210 & 9 & 15 & Antarctica \\
2028-06-03 05:47:31 & 6722865274261092224 & 16.21 & 842 & 23.2 & 43 & 101 & 212 & 6 & 11 & Chile, Argentina, Uruguay, S Brazil \\
2028-06-09 17:55:16 & 6722867301485782400 & 15.93 & 1322 & 24.2 & 41 & 102 & 213 & 5 & 10 & Namibia, South Africa \\
2028-07-18 01:46:24 & 6723049992215113728 & 16.19 & 6177 & 23.8 & 42 & 104 & 216 & 5 & 9 & none at night \\
2028-07-22 03:40:17 & 6723052397396893568 & 16.19 & 4807 & 23.2 & 43 & 104 & 217 & 5 & 10 & Antarctica \\
2028-07-24 06:02:37 & 6723052259958132608 & 15.07 & 856 & 22.8 & 44 & 104 & 217 & 5 & 10 & Patagonia \\
2029-05-13 17:49:39 & 6722558300044412800 & 12.26 & 2857 & 17.9 & 56 & 118 & 233 & 12 & 20 & Antarctica \\
2029-05-22 14:39:01 & 6722559850538626688 & 14.51 & 2201 & 20.4 & 49 & 118 & 233 & 10 & 17 & northern Australia \\
2029-06-22 23:38:59 & 6722942755466013440 & 14.93 & 3847 & 25.3 & 40 & 125 & 242 & 6 & 11 & Antarctica \\
2030-04-12 11:20:33 & 6722645200128753152 & 16.31 & 3266 & 8.3 & 120 & 270 & 428 & 27 & 44 & none at night \\
2030-04-28 12:01:43 & 6722643207263927936 & 13.87 & 5889 & 12.5 & 80 & 160 & 284 & 24 & 38 & Hawaii \\
\enddata
\tablecomments{Table~\ref{tab:occ} is published in its entirety (121 events with stars to $G\approx20.8$) in machine-readable format, which also gives the stellar position, Gaia $BP-RP$ color, renormalized unit weight error (RUWE; values $>1.4$ can indicate an unreliable astrometric solution, e.g., from an unresolved binary; \citealt{lindegren21}), expected magnitude drop, stellar position uncertainty projected at the distance of \ed, and the substellar point at closest approach; the 15 events with $G\le16.5$ are shown here. $b$: geocentric impact parameter of the shadow center; $v$: shadow velocity; $t_{1000}$: central duration for a diameter of 1000~km; $\sigma_\perp$ and $\sigma_t$: formal $1\sigma$ cross-track and timing uncertainties from the orbital covariance; $\sigma_\perp^{\rm sys}$ and $\sigma_t^{\rm sys}$: the same with an additional 20~mas error common to all CCD observations from the same night (Section~\ref{sec:3.2}). The last column lists land on the night side (star $>15^\circ$ above the horizon, Sun $<-12^\circ$) within $\pm1\sigma_\perp$ of the predicted path. Star positions include Gaia DR3 proper motions.}
\end{deluxetable*}

We propagated the orbital covariance to each event and projected it onto the shadow path, computing ground tracks on the World Geodetic System 1984 (WGS 84) reference ellipsoid \citep{nima00}. The formal $1\sigma$ cross-track uncertainty ($\sigma_\perp$ in Table~\ref{tab:occ}) is $\sim85-125$~km for most events in 2027-2029 ($\sim162$~km for the slow 2028 April 10~event), and $\sim160-270$~km in 2030, a fraction of the shadow width. The timing uncertainty is $\sim4-17$~s before 2030 and $\sim24-27$~s in 2030. The uncertainty in star positions also adds a further $\sim5-26$~km for these stars. These are thus formal uncertainties -- systematic errors here are more difficult to quantify. The Rubin residuals scatter by $\sim20$~mas in their nightly means, roughly double the formal uncertainty. We therefore repeated the calculations with an additional 20~mas error common to all CCD observations from the same night ($\sigma_\perp^{sys}$ and $\sigma_t^{sys}$ in Table~\ref{tab:occ}). This approximately doubles uncertainties, to $\sim190-290$~km in 2027-2029 and $\sim280-430$~km in 2030, and shifts the nominal orbit path by $\lesssim55$~km, except for the two slowest events, 2028 April 10 and 2030 April 12 ($\sim125$~km and $\sim165$~km). Adding instead a 0.1\arcsec\ floor to every observation, as in \Hcite, increases the cross-track uncertainties up to $\sim780$~km in 2028 and $\sim1010$~km in 2030, but this overestimates the measurement errors ($\chi^2 = 31$ for 220 degrees of freedom) and should be regarded as an upper limit. Between our adopted weighting and this floor, the nominal orbit path shifts by up to $\sim160$~km. Our predicted paths agree with those from the JPL orbit to within $\sim10-16$~km cross-track, with event times differing by $\lesssim10$~s. We caution that the true path uncertainties may reach a few hundred kilometers until Gaia DR4 and continued Rubin astrometry refine the orbit.

Because the cross-track uncertainties are much smaller than the predicted diameters of \ed, the James Webb Space Telescope (JWST) can also target occultations, which are not limited by weather or daylight. Repeating this same search with the current predicted JWST trajectory from JPL Horizons, we find 24 events in 2027-2030 with the shadow axis passing within $\sim2500$~km of the spacecraft whilst \ed\ is within the JWST field of regard. Over these events, we expect $\sim3-8$ positive detections for diameters $\sim635-1590$~km. The most favorable is the occultation of Gaia DR3 6722554421700899840 ($G=16.1$~mag) on 2029 April~17. On 2027 May~3, two stars (6722847029242779648, $G=19.7$~mag; 6722847029232687872, $G=20.3$~mag) are also occulted $\sim$81 minutes apart, with shadow-axis distances of $\sim724$ and $\sim253$~km from JWST, so a single visit could yield a chord and either a second one or a constraining non-detection. For most of the other events, the stars are faint ($G\sim20-21$~mag), and their Gaia DR3 proper-motion errors dominate the uncertainty -- Gaia DR4 will reduce these errors. These JWST predictions are provisional: the spacecraft's future position changes with station-keeping, and is reliable only a few months ahead \citep{santossanz16, proudfoot25}, so each event should be recomputed closer to its date.

\section{Photometric History and Shape} \label{sec:4}

\subsection{Phase Curve and Absolute Magnitude} \label{sec:4.1}
\begin{figure}
\centering
\includegraphics[width=\columnwidth]{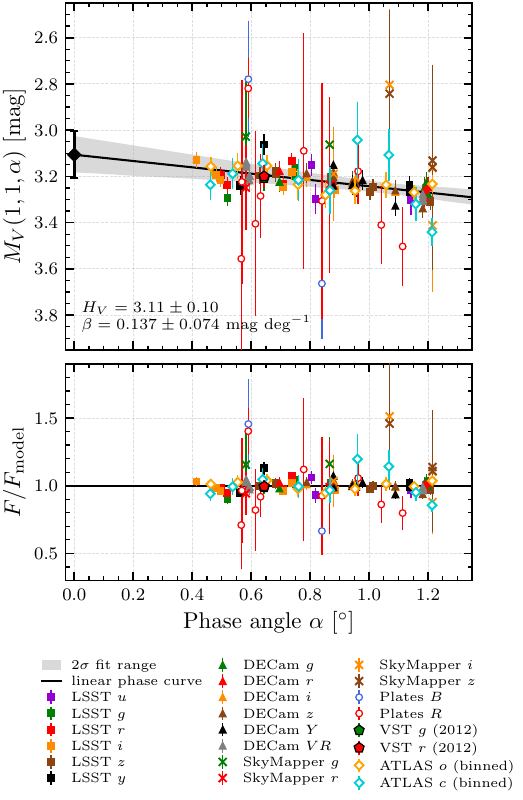}
\caption{Phase curve of \ed. Top: Distance reduced magnitude $M_V(1,1,\alpha)$ against phase angle $\alpha$. Every point is color-shifted to $V$ with the fitted band offsets. Colors mark the band and symbols the survey: Rubin LSST (squares), DECam (triangles) and SkyMapper (crosses; converted to the DECam bands) are fitted; VST 2012 (pentagons), ATLAS $o$ and $c$ (open diamonds; nightly means averaged in $0.1^\circ$ bins of phase angle and converted to $V$ with the colors of \ed) and photographic plates (open circles; 1954--2000) are overplotted but not fitted. Line and shading: linear fit and its 2$\sigma$ range; diamond: $H_V$ with 2$\sigma$ uncertainty; the $H_V$ and $\beta$ values printed in the panel are 2$\sigma$. Bottom: Flux relative to the linear phase curve model described in the main text. \label{fig:phase}}
\end{figure}

We fitted the 84 DECam, Rubin (in the Legacy Survey of Space and Time (LSST) filters), and SkyMapper magnitudes with a linear phase curve, $m_\lambda = H_\lambda + 5 \mathrm{log_{10}}(r_H \Delta) + \beta \alpha$, where $r_H$ and $\Delta$ are the heliocentric and geocentric distances, $\alpha$ is the Sun-target-observer phase angle, $\beta$ is a shared phase slope for all bands, and $H_\lambda$ is the absolute magnitude for a given band $\lambda$. We measured the SkyMapper magnitudes on difference images against the Dark Energy Camera Plane Survey 2 \citep[DECaPS2;][]{saydjari23}, calibrated them to Gaia DR3, and converted to the DECam bands so that they share their absolute magnitudes. Because the DECam and Rubin magnitudes are reported to the MPC without uncertainties, we adopted 0.03~mag for each (0.10~mag in $u$). We also included an additional intrinsic scatter term as a free parameter to account for rotational variability and other unmodeled effects; its best-fit maximum likelihood value is $\sim0.03$~mag. As seen in Figure~\ref{fig:phase}, the fit gives $\beta = 0.14 \pm 0.04$~mag~deg$^{-1}$, in agreement with \Hcite\, and $H_V = 3.11 \pm 0.05$~mag after converting $g$ and $r$ to $V$. The two VST magnitudes of 2012 and the ATLAS photometry of 2017-2026, which were not used in the fit, follow the same phase curve -- the VST $g$ and $r$ magnitudes lie within $\sim0.01$~mag of it.

\subsection{Seventy-two Years of Archival Photometry} \label{sec:4.3}
We measured plate magnitudes by fitting a model of the stellar image at the predicted position of \ed\ and calibrated against the Gaia DR3 magnitudes of the stars on the same plate. Due to differing techniques for plate development (emulsion and filter combinations, etc.), each one responds to a different range of wavelengths, so we included a color term in the calibration, evaluated at the color of \ed, and converted the result to Johnson $B$ or Cousins $R$ using colors computed from the solar spectrum multiplied by a linear reflectance slope of \ed. To compare the plates with modern data, we expressed each plate magnitude as an equivalent phase-corrected $V$ magnitude, with residuals given as $\Delta m$ in Table~\ref{tab:plates}. Their weighted mean is $\sim-0.02 \pm 0.06$~mag, with $\chi^2_\nu = 1.7$. This mean is only a summary; the shape analysis below fits each plate individually. Most of the excess comes from the two 1954 POSS-I plates, which lie $\sim0.50$~mag brighter than the later plates. The offset is also present in the independent SuperCOSMOS catalog photometry of the red plate, and we find no error in the epoch, filter, digitization, or calibration that accounts for it -- known asteroids on the 1954 plates match their JPL-predicted magnitudes. We cannot tell whether it reflects a calibration systematic of these plates or a real change in the brightness of \ed\ at that epoch, so we also report our shape constraints without these two plates.

This constancy constrains \ed's shape, as the viewing geometry changed substantially over the course of the archival observations -- between 1954 and 2026, the line of sight to \ed, measured in the object's frame, turned by $\sim87^\circ$. Unless the spin axis happens to lie in an unfavorable direction \citep[e.g., as in the case of (486958) Arrokoth;][]{spencer20}, a flattened body seen first more pole-on and later more equator-on (or the reverse) would have changed in brightness. To quantify this, we modeled \ed\ as an ellipsoid with semi-axes $a\ge b\ge c$, whose brightness at the small phase angles of all of our observations ($\alpha \le1.2^\circ$) is proportional to its projected area for a Lommel-Seeliger scattering surface \citep{muinonen15}. For an oblate spheroid ($a=b$), we computed the predicted change in brightness between each plate epoch and the mean modern geometry for 3000 spin-axis directions spread uniformly over the sphere. The uncertainty model adds a 0.2~mag per-plate systematic and a common zero-point term of 0.03~mag, and, for triaxial shapes, adds the variance due to the unknown rotational phase of every single plate. For each combination of shape and spin axis, we compared the resulting $\chi^2$ with that of a body of constant brightness, and excluded the combination if its $\chi^2$ exceeded that of constant brightness by more than 3.84, the 95\% confidence threshold for one parameter. Shapes that fit better than constant brightness are never excluded, so this test is conservative.

Oblate shapes with $c/a\le0.66$ are excluded for at least half of all directions, and those with $c/a\le0.54$ for 68\%. Without the two 1954 plates, shapes with $c/a\leq0.48$ are still excluded for at least half of all directions. No shape is excluded for all directions, because a spin-axis near the great circle midway between the 1954 and 2026 lines of sight presents the same aspect at both epochs. Treating the spin-axis direction as unknown and isotropic, and $c/a$ as uniformly distributed, the marginal constraint is $c/a>0.38$ at 95\% confidence. For scale, at $c/a=0.7$, the predicted change between 1954 and today reaches $\sim0.39$~mag, with a median of $\sim0.2$~mag over spin-axis directions, whereas at $c/a=0.9$ it never exceeds $\sim0.11$~mag, below our sensitivity given our per-plate uncertainties of $\sim0.24-0.55$~mag. Such a change is not excluded for all spin-axis orientations because its sign depends on the orientation. The plate photometry is consistent with constant brightness, and thus with a spherical or nearly-spherical body, although it does not by itself require one.

\subsection{Rotational Light Curve} \label{sec:4.4}
A rotating triaxial body varies in brightness even at fixed viewing geometry, so a measurement of the light curve amplitude constrains $b/a$. ATLAS observed \ed\ on 401 nights from 2017-2026, with single 30~s frames reaching S/N$\sim2$ in crowded fields. We therefore performed forced photometry on difference images to improve S/N, with photometry calibrated again to Gaia. We rejected frames with a star more than three times brighter than \ed\ at its position, yielding 1638 possible usable frames that agree with the phase curve to $0.001 \pm 0.012$~mag in $o$.

\begin{figure}
\centering
\includegraphics[width=\columnwidth]{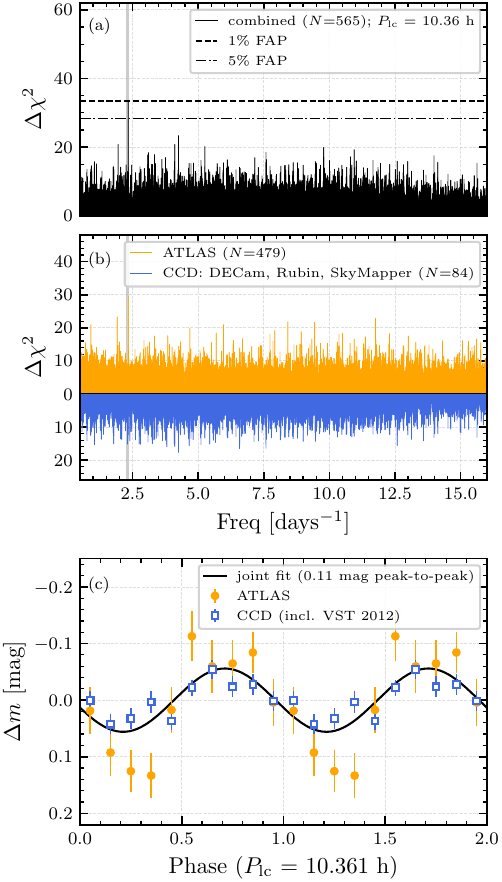}
\caption{Rotation search on the combined sparse photometry of \ed. (a) $\Delta\chi^2$ periodogram for a sinusoid plus the phase-curve model with per-night correlated noise. The fit uses the ATLAS visits and the DECam, Rubin, SkyMapper, and VST magnitudes. Dashed and dash-dotted lines: 1\% and 5\% false-alarm levels from 200 noise simulations. Grey band: $P_{\rm lc}=10.36$\,h. (b) Similar to (a), but for ATLAS alone (upper) and CCD data alone (lower). (c) Residuals folded at $P_{\rm lc}$ and averaged in phase bins for ATLAS (filled circles) and the CCD data (open squares), with the joint sinusoid (0.11\,mag peak-to-peak). Two cycles are shown. \label{fig:rot}}
\end{figure}

For the rotation search, we averaged the frames of each ATLAS quad (479 quads of up to four 30~s frames taken within about an hour). We fitted these jointly with the 86 DECam, Rubin, SkyMapper, and VST magnitudes, using our phase curve model from Section~\ref{sec:4.1} plus a sinusoid, via a generalized least squares periodogram \citep{lomb76, scargle82, zechmeister09}. Our strongest peak as seen in Figure~\ref{fig:rot} lies at $P_{\rm lc} = 10.36$~h, with a peak-to-peak amplitude of 0.11~mag, and a false-alarm probability (FAP) of 1.5\%. At this period, the CCD data alone show a signal with the same phase at $\sim3.3\sigma$ (FAP $5\times10^{-4}$, amplitude $0.08 \pm 0.02$~mag). The phase is consistent across Rubin, DECam, 2017-2025 ATLAS, and 2026 ATLAS, although around half of the signal comes from 2026. We therefore regard 10.4~h (single-peaked) or 20.7~h (double-peaked) as the candidate rotation period. Aliases at 7.2 and 18.2~h, one cycle per day away, are not excluded, and denser time-series photometry will be needed for clarification. Our photographic plates do not change this result.

We also injected double-peaked light curves with periods in the range of 4-30~h. Peak-to-peak amplitudes of 0.10, 0.15, and 0.20~mag are recovered in $\sim42\%$, $\sim95\%$, and $\sim100\%$ of trials, so a 0.1~mag light curve here lies at the edge of detectability. For our shape analysis, we therefore adopt two amplitude thresholds: 0.11~mag, which attributes all DECam and Rubin night-to-night scatter to rotation, and 0.15~mag, above which 95\% of injected light curves are recovered. With the plate photometry, these limits exclude a Haumea-like Jacobi ellipsoid \citep[$b/a\sim0.8$;][]{chandrasekhar69, ortiz17} for $\sim88-94$\% of spin-axis directions. We also note that this amplitude is consistent with the distribution of light-curve amplitudes for $H_r < 6$ objects measured by \cite{bernardinelli23}, consistent with what is expected if this object is relatively round.

Whilst most spin-axis directions disfavor strong flattening, they do not test for hydrostatic equilibrium. A Maclaurin spheroid \citep{chandrasekhar69} with $\rho = 1.55$~g~cm$^{-3}$ and a period longer than $\sim6-8$~h has $c/a\gtrsim0.7-0.85$, which could produce a brightness change below our sensitivity. The candidate 10.4~h period, if confirmed, would place \ed\ in this slowly rotating regime -- however, bodies at the low end of the size range of \ed\ need not be relaxed \citep{tancredi08}. An occultation would allow us to measure the shape of \ed\ more directly.

\section{Surface Composition} \label{sec:5}

\subsection{Near-infrared Colors and Water Ice}
\begin{figure*}
\centering
\includegraphics[width=\textwidth]{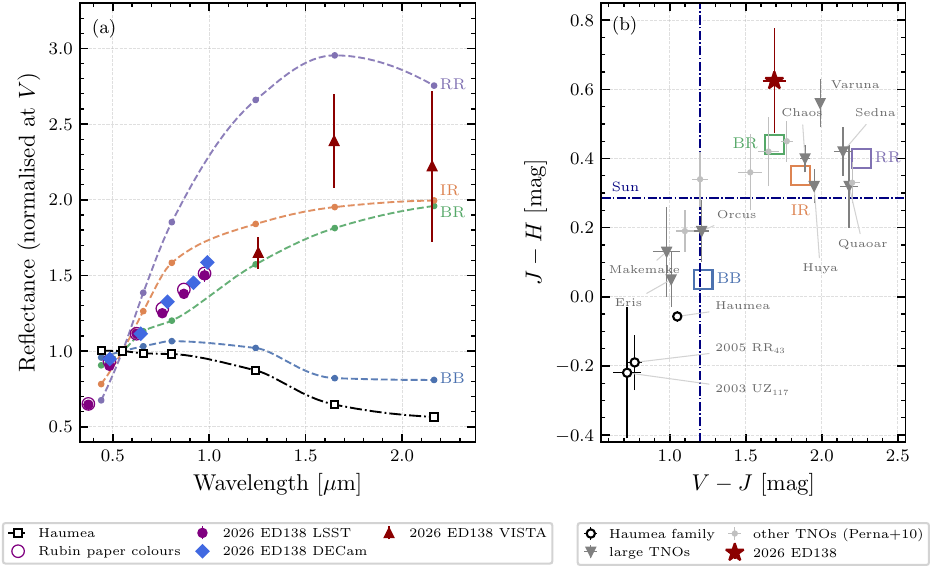}
\caption{Surface colors of \ed. (a) Broadband reflectance relative to the Sun, normalized at $V$, from Rubin (LSST) and DECam per-band absolute magnitudes and VISTA $J$, $H$ and $K_s$ magnitudes corrected to the geometry of the Rubin observations with the Section~\ref{sec:4.1} phase curve and placed on the same scale through their colors relative to Rubin $r$; open circles are the colors of \Hcite. Dashed curves: BB, BR, IR and RR class mean colors \citep{barucci05,fulchignoni08} at the band centers (dots), joined by cubic spline interpolation for display; they are not fits to \ed. Dash-dotted with squares: colors of Haumea for comparison. (b) $J-H$ against $V-J$, with \ed\ (star) transformed to the 2MASS-like scale of \citet{gonzalezfernandez18}. Open circles: Haumea \citep{lacerda09} and family members 2005~RR$_{43}$ and 2003~UZ$_{117}$ \citep{perna10}. Triangles: large TNOs. Dots: other TNOs \citep{perna10}. Open squares: class means. Dash-dotted lines: solar colors. \label{fig:sed}}
\end{figure*}

On 2010 March~18, VISTA observed \ed\ in $J$, $H$, and $K_s$ within 17~min, short enough that rotation should not affect the measured colors of $J-H = 0.60\pm0.15$ and $H-K_s = 0.01 \pm0.27$. Two further $J$ epochs, 2010 March~25 and 2017 July~22, agree with the first after correcting for viewing geometry, and with the phase curve give $V-J = 1.71\pm0.08$ (VISTA system, corresponding to 1.69 on the 2MASS-like scale of Figure~\ref{fig:sed}). We converted the colors to relative reflectance using the solar colors of \citet{willmer18}. The broadband reflectance rises from the visible, with a slope of $\sim12\%$~per~100~nm from $g$ to $y$ to $H$. The $J$ point lies $\sim1.8\sigma$ below the extrapolated visible slope, which is consistent with the near-infrared flattening common among red TNOs seen in Figure~\ref{fig:sed} \citep{hainaut12, fraser12, seccull21}. Compared with the class means and dispersions of \citet{barucci05} and \citet{fulchignoni08} in $B-V$, $V-R$, $V-I$, $V-J$, $V-H$, and $V-K$, \ed\ falls in the BR (intermediate red) class, and no other class lies within 3$\sigma$.

Water ice has strong absorption bands at 1.5 and 2.0~$\mu$m, which lower the $H$ flux relative to $J$, giving icy surfaces blue $J-H$ colors, as seen for Haumea and its collisional family \citep{brown07b, snodgrass10}. However, the $J-H$ color of \ed\ is redder than both that of Haumea \citep[$-0.04$ to $-0.06$;][]{trujillo07, lacerda09b} and of family member 2005~RR$_{43}$ \citep{perna10} by $\sim4.5\sigma$. To turn this into a naive ice limit, we modeled a surface as an areal mixture of a Haumea-like icy component and a red, ice-free continuum. We fitted the mixing fraction to $J-H$ and $H-K_s$ only, so that the visible slope cannot mimic ice. The best fit has no icy component, and at $\sim95\%$ confidence that component contributes at most $\sim30-45\%$ of the $J$ flux, depending on the assumed continuum. Because ice is much brighter than a dark red surface, in this two-component model, this corresponds to $\lesssim15\%$ of the surface area for $p_V\sim0.15$ ($\sim7-25\%$ for $p_V=0.08-0.25$). This independently supports the conclusion of \Hcite\, from the visible colors and dynamics, that \ed\ is not a Haumea-family member.

This test, however, rests on a single $H$-band measurement whose zero point is set by only six stars. Nonetheless, a Haumea-like $J-H$ would require an error of $\sim0.6$~mag in $H$-band, against its uncertainty of 0.13~mag ($\sigma_{J-H} = 0.15$). To test whether field stars could have biased the $H$ measurement, we checked for contamination from any bright stars within 4\arcsec\ of \ed; all difference images showed clean point sources in both $J$ and $H$. Measuring flux on a ring around the nearest star to \ed\ at the same distance amounts to only a $\sim3\%$ and $\sim9\%$ reduction in flux of \ed\ in $J$ and $H$ respectively, which could redden $J-H$ by $\sim0.05$~mag, but would not make it substantially bluer. Additional near-infrared measurements would nevertheless allow for its colors to be confirmed.

\subsection{Methane Ice} \label{5.2}
Methane ice, which dominates the spectra of Eris \citep{brown05, licandro06} and Makemake \citep{licandro06b, brown07}, has a strong absorption band at 0.89~$\mu$m \citep{grundy02}, within the $z$ band. We tested for this photometrically by fitting a linear reflectance continuum to the $g-r$ and $r-i$ colors, integrating it under the bandpasses to predict the $z$ magnitude, and comparing our residuals, assuming that positive $\Delta z$ implies absorption. For \ed, the values of $\Delta z = -0.005 \pm 0.024$~mag for DECam and $0.055\pm0.030$~mag for Rubin are consistent with the continuum. Eris, measured the same way, gives $\Delta z = 0.162\pm0.028$~mag. This index is not specific to methane because near-infrared flattening would also make $\Delta z$ positive, but a null result here still sets an upper limit on methane absorption. 

To quantify this limit, we modeled a surface where methane ice covers some fraction $f$, with the remainder matching the continuum. In the ice-covered fraction, the continuum is dimmed by $e^{-\alpha_{abs}L}$, where $\alpha_{abs}$ is the laboratory absorption coefficient of methane ice \citep{grundy02, SSHADE_CH4} and $L$ is the effective path length of light through the ice. We integrated this model over the bandpasses and fitted it to $\Delta z$ and the analogous $Y$-band index $\Delta Y$, with their covariance from Monte Carlo resampling of the colors. For a fully covered surface ($f=1$), the combined DECam and Rubin data give $L < 0.83$~cm, $\sim20\%$ of the $\sim3.9$~cm we derive for Eris in the same way. Fixing $L$ at the Eris value instead, and assuming that the ice-free fraction shares the continuum reflectance, ice as absorbing as that on Eris can cover at most $\sim25$\% of the surface. 2005 RN$_{43}$, however, whose near-infrared spectrum shows no ice absorption \citep{brown12}, displays $\Delta z \sim +0.1$ from near-infrared flattening alone, whereas Gonggong, which has at most trace methane absorption and a very red spectrum \citep{brown11, emery24}, has $\Delta z \sim -0.07$. This index can therefore be biased for very red objects, but the colors of \ed\ indicate only a moderately red surface, so we do not expect this to affect our limits. Much weaker Quaoar-like methane absorption \citep{schaller07, fraser13} is below our detection limit. The absence of strong methane agrees with the prediction of \Hcite\ that \ed\ is too small and warm to retain surface methane. With no evidence for Haumea-like water ice, this makes a very bright ($p_V\gtrsim0.5$) icy surface, and so a diameter below $\sim450$~km, less likely, although broadband colors cannot exclude a bright but spectrally bland surface. 

\section{Long-term Dynamics} \label{sec:6}
\Hcite\ found that \ed\ librates in the mixed 12:7 resonant argument $\phi=12\lambda - 7\lambda_N - 3\varpi - 2\Omega$ (where $\lambda$, $\varpi$, and $\Omega$ are the mean longitude, longitude of perihelion, and longitude of the ascending node of \ed, and $\lambda_N$ is the mean longitude of Neptune), but that it lies close to the edge of resonance. TNOs can alternate between resonant and non-resonant behavior in a process known as resonance sticking \citep{lykawka_exploring_2006, lykawka_resonance_2007}, which occurred for about half of the \Hcite~clones over their 10~Myr integration period. To determine how \ed\ behaves over much longer times, we integrated clones of its orbit for 1~Gyr forwards and backwards in time.

We drew 200 clones from the covariance of the adopted orbit in Table~\ref{tab:fit}, and another 100 from the covariance scaled by a factor of nine (i.e., uncertainties increased threefold) to allow for systematic errors. We integrated them in \texttt{REBOUND} as massless particles with the Sun (including the mass of the terrestrial planets) and the four giant planets, using the WHFast integrator \citep{wisdom_symplectic_1991, rein_whfast_2015} with a timestep of 180~days. We sampled $\phi$ every 250~yr and, in successive 2~Myr windows, classified a clone as librating if $\phi$ did not circulate, i.e., if its deviation from the mean value in that window never exceeded 175$^\circ$. This criterion allows us to distinguish libration from circulation, rather than imposing a maximum libration amplitude. A stricter threshold of 150$^\circ$ reduces the median duration of continuous forward libration to $\sim2-4$~Myr, whilst the time-averaged librating fractions change by $<0.1$. The stricter threshold shortens forward durations as the present $\sim125^\circ$ amplitude grows to $\sim150^\circ$ within a few megayears. We removed clones whose orbits became hyperbolic or reached $a>200$~au.

\begin{figure}
\centering
\includegraphics[width=\columnwidth]{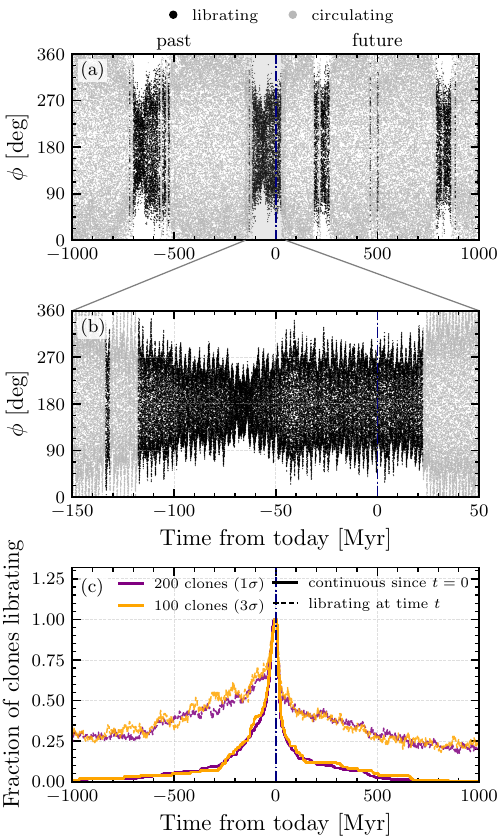}
\caption{Resonance sticking of \ed\ over $\pm1$\,Gyr. (a) Mixed 12:7 resonant argument $\phi=12\lambda-7\lambda_N-3\varpi-2\Omega$ of the nominal orbit, sampled every 25~kyr; black where $\phi$ librates and gray where it circulates. The shaded interval is enlarged in (b), sampled every 2~kyr. (c) Fraction of clones librating, for 200 clones drawn from the orbital covariance (purple) and 100 from the covariance scaled by nine (orange). Solid: fraction librating continuously since today; dashed: fraction librating at time $t$ (maximum deviation of $\phi$ from its center $<175^\circ$ in a 2~Myr window). After 1~Gyr, 20--30\% of clones are librating at any one time. \label{fig:dyn}}
\end{figure}

Figure~\ref{fig:dyn} shows that all clones librate today, about a center near $180^\circ$, and with a maximum excursion of $\sim125^\circ$ per window, consistent with the libration amplitude of $\sim95-115^\circ$ that \Hcite\ measured over shorter, 200~kyr windows. Over the next 10~Myr, $\sim78-86$\% of the clones librate continuously by our criterion, compared with $\sim46\%$ using the stricter \citet{khain20} criterion adopted in \Hcite. Over longer times, the libration becomes intermittent -- continuous libration lasts a median of $\sim16-20$~Myr forward in time, and $\sim42-49$~Myr backwards, and only $\sim1\%$ of clones librate for the full gigayear. Averaged over 1~Gyr, the clones spend $\sim32-33\%$ of their time librating in the forward integrations, and $\sim41-43\%$ in the backwards ones. Within 1~Gyr, $\sim10-20\%$ of clones are removed, and $\sim20-38\%$ reach perihelion distances $q<33$~au, where close encounters with Neptune become possible.

Because the motion is chaotic, the backwards integrations do not reconstruct the actual history of \ed\ -- rather, they show the range of histories consistent with its present orbit. In both directions, the behavior is the same: \ed\ repeatedly enters and leaves the 12:7 resonance, sticking to it for tens of megayears at a time, and is not permanently captured. Its present resonant state is therefore probably temporary.

\section{Summary} \label{sec:7}

\ed\ was recorded on photographic plates for more than half a century before Rubin Observatory discovered it. The earliest confirmed detection is from 1954, with a marginal detection in 1937. We also identified \ed\ in hundreds of CCD images from 2010 onwards. The resulting 89~yr arc changes little for expectations of 2027 occultations, whose paths are largely fixed by the 2026 Rubin data, but it improves the long-term ephemeris fivefold. Over 72~yr, whilst our viewing direction turned by $\sim87^\circ$, the plate magnitudes of \ed\ remain consistent with the present-day phase curve, which disfavors strong flattening for most spin-axis orientations. Combined with ATLAS and other CCD photometry, we obtain a candidate rotation period of 10.4~h (single-peaked) or 20.7~h (double-peaked) with an amplitude of $\sim0.1$~mag. VISTA near-infrared colors hint towards a BR-class surface without the 1.5~$\mu$m water-ice feature of the Haumea family, and $z$-band photometry shows no indication of strong methane absorption. Dynamically, \ed\ sticks to the 12:7 resonance with Neptune for tens of megayears at a time, but it is not permanently captured.

The archival observations of \ed\ in this paper illustrate how much scientific value existing data can add to a newly discovered solar system object. This is particularly true near the galactic plane, where crowding has hampered moving-object searches, but where photographic and near-infrared survey coverage is extensive. The approach outlined in this work could also apply to returning comets, recurrently active Centaurs, and the hypothesized Planet Nine \citep[e.g.,][]{batygin16}. Much of Planet Nine's remaining unsearched parameter space lies near the galactic plane \citep{holman25b}, and its predicted brightness of $V = 22.0^{+1.1}_{-1.4}$~mag \citep{brown24} places the bright end of its range within the reach of single Schmidt plates. As it would move only a few degrees across the photographic era, stacking plates along trial orbits could theoretically reach fainter candidates. As Rubin continues to find additional objects in such regions, a systematic search of photographic and CCD archives along their orbits can, within days to weeks of discovery, extend their arcs by decades, provide new multi-wavelength color measurements, and constrain their shapes by exploiting viewing geometries that will not recur on timescales of a century or more.

\begin{acknowledgments}
J.M., C.O.C., M.J., I.C., and R.S. acknowledge support from the DiRAC Institute in the Department of Astronomy at the University of Washington College of Arts and Sciences. The DiRAC Institute is supported through generous gifts from the Charles and Lisa Simonyi Fund for Arts and Sciences, Janet and Lloyd Frink, and the Washington Research Foundation. 

C.O.C. is supported by Schmidt Sciences. C.O.C. gratefully acknowledges support from the NASA CSSFP (grant No. 80NSSC26K0380).

I.C. acknowledges the support of the Natural Sciences and Engineering Research Council of Canada (NSERC) and is partially funded by an NSERC Postgraduate Scholarship-Doctoral. 

P.H.B. and B.M.P. acknowledge support from São Paulo Research Foundation (FAPESP), grant number 2026/16152-4.

M.J.H. and K.J.N. acknowledge support from the NSF (grant No. AST-2206194) and the NASA YORPD Program (grant No. 80NSSC22K0239). M.J.H. acknowledges support from the LSST Discovery Alliance (LSST-DA) through LINCC Frameworks Incubator grant 2025-SFF-LFI-10-Holman.

This work was carried out as part of the ongoing Small Body Search and Rescue (\href{https://sbsar.net/}{SBSAR}) project for understudied small solar system objects, co-led by C.O.C. and J.M.

This research has made use of data and/or services provided by the International Astronomical Union's Minor Planet Center. This research has made use of NASA's Astrophysics Data System Bibliographic Services. This research has made use of the Science Explorer, funded by NASA under Cooperative Agreement 80NSSC21M00561. 

The Digitized Sky Surveys were produced at the Space Telescope Science Institute under U.S. Government grant NAG W-2166. The images of these surveys are based on photographic data obtained using the Oschin Schmidt Telescope on Palomar Mountain and the UK Schmidt Telescope. The plates were processed into the present compressed digital form with the permission of these institutions. The National Geographic Society - Palomar Observatory Sky Atlas (POSS-I) was made by the California Institute of Technology with grants from the National Geographic Society. The UK Schmidt Telescope was operated by the Royal Observatory Edinburgh, with funding from the UK Science and Engineering Research Council (later the UK Particle Physics and Astronomy Research Council), until 1988 June, and thereafter by the Anglo-Australian Observatory. The blue plates of the southern Sky Atlas and its Equatorial Extension (together known as the SERC-J), as well as the Equatorial Red and the Second Epoch [red] Survey were all taken with the UK Schmidt.

This research has made use of data obtained from the SuperCOSMOS Science Archive, prepared and hosted by the Wide Field Astronomy Unit, Institute for Astronomy, University of Edinburgh, which is funded by the UK Science and Technology Facilities Council. We used plates from the ESO 1-m Schmidt Telescope at La Silla and the UK Schmidt Telescope, including the AAO/UKST SuperCOSMOS H$\alpha$ Survey.

This work has made use of data provided by Digital Access to a Sky Century $@$ Harvard (DASCH), which has been partially supported by NSF grants AST-0407380, AST-0909073, and AST-1313370. Work on DASCH Data Release 7 received support from the Smithsonian American Women's History Initiative Pool.

Funding for APPLAUSE has been provided by DFG (German Research Foundation, Grants EN926/3-1/2, HE1356/63-1/2, GR969/4-1, SCHM1032/552), the Leibniz Institute for Astrophysics Potsdam, Dr.~Remeis Sternwarte Bamberg (University N\"urnberg/Erlangen), the Hamburger Sternwarte (University of Hamburg) and Tartu Observatory.

Part of this work was based on observations collected at the European Southern Observatory under ESO programmes 179.B-2002, 198.B-2004 and 177.D-3023.

The national facility capability for SkyMapper has been funded through ARC LIEF grant LE130100104 from the Australian Research Council, awarded to the University of Sydney, the Australian National University (ANU), Swinburne University of Technology, the University of Queensland, the University of Western Australia, the University of Melbourne, Curtin University of Technology, Monash University and the Australian Astronomical Observatory. SkyMapper is owned and operated by the ANU's Research School of Astronomy and Astrophysics. The survey data were processed and provided by the SkyMapper Team at ANU. The SkyMapper node of the All-Sky Virtual Observatory (ASVO) is hosted at the National Computational Infrastructure. Development and support of the SkyMapper node of the ASVO has been funded in part by Astronomy Australia Limited and the Australian Government through the Commonwealth's Education Investment Fund and National Collaborative Research Infrastructure Strategy, particularly the National eResearch Collaboration Tools and Resources and the Australian National Data Service Projects.

This work has made use of data from the Asteroid Terrestrial-impact Last Alert System (ATLAS) project. ATLAS is primarily funded to search for near-Earth asteroids through NASA grants NN12AR55G, 80NSSC18K0284, 80NSSC18K1575, and 80NSSC18K0265; byproducts of the NEO search include images and catalogs from the survey area. The ATLAS science products have been made possible through the contributions of the University of Hawaii Institute for Astronomy, the Queen's University Belfast, the Space Telescope Science Institute, the South African Astronomical Observatory, and the Millennium Institute of Astrophysics, Chile.

This material is based upon work supported in part by the National Science Foundation through Cooperative Agreements AST-1258333 and AST-2241526 and Cooperative Support Agreements AST-1202910 and 2211468 managed by the Association of Universities for Research in Astronomy (AURA), and the Department of Energy under Contract No.~DE-AC02-76SF00515 with the SLAC National Accelerator Laboratory managed by Stanford University. Additional Rubin Observatory funding comes from private donations, grants to universities, and in-kind support from LSST-DA Institutional Members.

This project used data obtained with the Dark Energy Camera (DECam), which was constructed by the Dark Energy Survey (DES) collaboration. Funding for the DES Projects has been provided by the DOE and NSF (USA), MISE (Spain), STFC (UK), HEFCE (UK), NCSA (UIUC), KICP (U.~Chicago), CCAPP (Ohio State), MIFPA (Texas A\&M), CNPQ, FAPERJ, FINEP (Brazil), MINECO (Spain), DFG (Germany) and the Collaborating Institutions in the Dark Energy Survey, which are Argonne Lab, UC Santa Cruz, University of Cambridge, CIEMAT-Madrid, University of Chicago, University College London, DES-Brazil Consortium, University of Edinburgh, ETH Z\"urich, Fermilab, University of Illinois, ICE (IEEC-CSIC), IFAE Barcelona, Lawrence Berkeley Lab, LMU M\"unchen and the associated Excellence Cluster Universe, University of Michigan, NSF NOIRLab, University of Nottingham, Ohio State University, OzDES Membership Consortium, University of Pennsylvania, University of Portsmouth, SLAC National Lab, Stanford University, University of Sussex, and Texas A\&M University. Based on observations at NSF Cerro Tololo Inter-American Observatory, NSF NOIRLab (NOIRLab Prop. IDs 2014A-0429, 2016A-0327, 2016B-0279, 2018A-0251, 2018B-0271 and 2019A-0265; PI: D. Finkbeiner), which is managed by AURA under a cooperative agreement with the U.S.\ National Science Foundation.

The Legacy Surveys consist of three individual and complementary projects: the Dark Energy Camera Legacy Survey (DECaLS; Proposal ID \#2014B-0404; PIs: David Schlegel and Arjun Dey), the Beijing-Arizona Sky Survey (BASS; NOAO Prop.\ ID \#2015A-0801; PIs: Zhou Xu and Xiaohui Fan), and the Mayall z-band Legacy Survey (MzLS; Prop.\ ID \#2016A-0453; PI: Arjun Dey). DECaLS, BASS and MzLS together include data obtained, respectively, at the Blanco telescope, Cerro Tololo Inter-American Observatory, NSF's NOIRLab; the Bok telescope, Steward Observatory, University of Arizona; and the Mayall telescope, Kitt Peak National Observatory, NOIRLab. Pipeline processing and analyses of the data were supported by NOIRLab and the Lawrence Berkeley National Laboratory (LBNL). The Legacy Surveys project is honored to be permitted to conduct astronomical research on I'oligam Du'ag (Kitt Peak), a mountain with particular significance to the Tohono O'odham Nation. NOIRLab is operated by AURA under a cooperative agreement with the National Science Foundation. LBNL is managed by the Regents of the University of California under contract to the U.S.\ Department of Energy. BASS is a key project of the Telescope Access Program, which has been funded by the National Astronomical Observatories of China, the Chinese Academy of Sciences (the Strategic Priority Research Program ``The Emergence of Cosmological Structures'' Grant \# XDB09000000), and the Special Fund for Astronomy from the Ministry of Finance. The BASS is also supported by the External Cooperation Program of Chinese Academy of Sciences (Grant \# 114A11KYSB20160057), and Chinese National Natural Science Foundation (Grant \# 12120101003, \# 11433005). The Legacy Survey team makes use of data products from the Near-Earth Object Wide-field Infrared Survey Explorer (NEOWISE), which is a project of the Jet Propulsion Laboratory/California Institute of Technology. NEOWISE is funded by the National Aeronautics and Space Administration. The Legacy Surveys imaging of the DESI footprint is supported by the Director, Office of Science, Office of High Energy Physics of the U.S.\ Department of Energy under Contract No.~DE-AC02-05CH11231, by the National Energy Research Scientific Computing Center, a DOE Office of Science User Facility under the same contract; and by the U.S.\ National Science Foundation, Division of Astronomical Sciences under Contract No.~AST-0950945 to NOAO.

This work has made use of data from the European Space Agency mission {\it Gaia} (\url{https://www.cosmos.esa.int/gaia}), processed by the {\it Gaia} Data Processing and Analysis Consortium (DPAC, \url{https://www.cosmos.esa.int/web/gaia/dpac/consortium}). Funding for the DPAC has been provided by national institutions, in particular the institutions participating in the {\it Gaia} Multilateral Agreement.

This publication makes use of data products from the Two Micron All Sky Survey, which is a joint project of the University of Massachusetts and the Infrared Processing and Analysis Center/California Institute of Technology, funded by the National Aeronautics and Space Administration and the National Science Foundation.

This research used the facilities of the Canadian Astronomy Data Centre operated by the National Research Council of Canada with the support of the Canadian Space Agency. This research has made use of the CATCH tool and data archives of the NASA Planetary Data System Small Bodies Node and of the JPL Horizons system. Laboratory absorption coefficients of methane ice were obtained from the SSHADE database (\url{https://www.sshade.eu}).

Portions of the code used to query archives and produce cutouts and some analysis for this work were written with the assistance of large-language-model coding tools (Anthropic \texttt{Claude Code}, Opus 4.8, Opus 5.5, and Fable 5.1 models; OpenAI Codex, Astra-6 and Luna-6 models). The authors ran, checked, and interpreted all code and verified every reported observation by eye, rejecting any false detections. The same tools were used to assist with refining, spell-checking, and organizing the manuscript text; Grammarly was also used. The authors have reviewed and approved all scientific content, interpretations, and conclusions in this work, and take full responsibility for the final manuscript.

\end{acknowledgments}

\begin{contribution}
J.M. led the study's conceptualization, methodology, investigation, analysis, data scraping and curation, visualization, and preparation of the manuscript. 

C.O.C. led the development of the code that was used to search and download images of solar system objects. He also provided guidance on analysis and manuscript writing, as well as project direction.

M.J. provided guidance on project direction, including suggestions of older plate searches, as well as manuscript preparation advice.

I.C. provided detailed feedback on the manuscript, including review and editing, which greatly improved the quality of the work.

P.H.B. provided review and edits of the manuscript, as well as discussion on directions for analysis and interpretation of color and rotation results.

B.M.P. provided a cross-check on the rotational light curve analysis.

R.S. provided feedback and interpretation of the occultation predictions.

S.D. and H.T. both independently provided a cross-check on each of our photographic plate measurements to ensure they were valid detections.

M.J.H. and K.J.N. are both responsible for the initial detection of the dwarf planet candidate within Rubin data, and also provided archival DECam observations back to 2016 prior to this work.


\end{contribution}

%
\facilities{PO:1.2m, UKST, ESO:Schmidt, ESO:VISTA (VIRCAM), VST (OmegaCAM), SkyMapper, Blanco (DECam), Rubin:Simonyi (LSSTCam), ATLAS, Gaia, CTIO:2MASS, MPC, PDS}

\software{
\texttt{astrometry.net} \citep{lang10},
\texttt{astropy} \citep{astropy13,astropy18,astropy22},
\texttt{astroquery} \citep{ginsburg19},
\texttt{photutils} \citep{photutils},
\texttt{reproject} \citep{reproject},
\texttt{NumPy} \citep{harris20},
\texttt{SciPy} \citep{virtanen20},
\texttt{Matplotlib} \citep{hunter07},
\texttt{pandas} \citep{mckinney10, pandas20},
\texttt{layup} \citep{holman26b},
\texttt{ASSIST} \citep{holman23,rein23},
\texttt{REBOUND} \citep{rein12,rein15},
\texttt{Find\_Orb} \citep{findorb},
\texttt{astrochipmunk}\footnote{In development by C.O.C.; based in part on methods described in \cite{chandlerChasingTailsActive2022a}.}
}

\bibliography{zrefs}{}
\bibliographystyle{aasjournalv7.1}

\end{document}